\documentclass[runningheads]{llncs}
\usepackage[T1]{fontenc}
\usepackage{graphicx}
\usepackage{hyperref}
\begin{document}
\title{Enabling Cyber Security Education through Digital Twins and Generative AI}
%
%
\author{Vita Santa Barletta\inst{1} \and
Vito Bavaro\inst{1} \and
Miriana Calvano\inst{1} \and
Antonio Curci\inst{1} \and
Antonio Piccinno\inst{1} \and
Davide Pio Posa \inst{1}
}
\authorrunning{Barletta et al.}
%
\institute{University of Bari Aldo Moro
\email{\{vita.barletta,miriana.calvano,antonio.curci, antonio.piccinno\}@uniba.it}\\\{v.bavaro4,d.posa3\}@studenti.uniba.it}
\maketitle              
\begin{abstract}
Digital Twins (DTs) are gaining prominence in cybersecurity for their ability to replicate complex IT (Information Technology), OT (Operational Technology), and IoT (Internet of Things) infrastructures, allowing for real-time monitoring, threat analysis, and system simulation. This study investigates how integrating DTs with penetration testing tools and Large Language Models (LLMs) can enhance cybersecurity education and operational readiness. By simulating realistic cyber environments, this approach offers a practical, interactive framework for exploring vulnerabilities and defensive strategies.
At the core of this research is the Red Team Knife (RTK), a custom penetration testing toolkit aligned with the Cyber Kill Chain model. RTK is designed to guide learners through key phases of cyber-attacks including reconnaissance, exploitation, and response—within a DT-powered ecosystem. The incorporation of Large Language Models (LLMs) further enriches the experience by providing intelligent, real-time feedback, natural language threat explanations, and adaptive learning support during training exercises.
This combined DT–LLM framework is currently being piloted in academic settings to develop hands-on skills in vulnerability assessment, threat detection, and security operations. Initial findings suggest that the integration significantly improves the effectiveness and relevance of cybersecurity training, bridging the gap between theoretical knowledge and real-world application. Ultimately, the research demonstrates how DTs and LLMs together can transform cybersecurity education to meet evolving industry demands.
 
\end{abstract}
\section{Introduction} \label{sec:intro}
Traditional cybersecurity has historically focused on protecting technical infrastructures, systems, and computer networks. However, the evolution of the threat landscape has highlighted how the human element often represents the weakest link in the security chain. Therefore, Cyber Social Security emerges as a discipline that studies the intersection between social factors, human behaviors, and information security, recognizing that many of the most sophisticated attacks exploit psychological and social vulnerabilities rather than exclusively technical ones. 
The advent of Generative Artificial Intelligence has revolutionized this scenario, significantly amplifying attackers' capabilities to conduct large-scale social engineering operations. In this context, Digital Twins (DTs) are increasingly recognized as powerful assets in cybersecurity, offering real-time monitoring, detailed system analysis, and high-fidelity emulation capabilities \cite{Luzzi_2024}. By creating accurate virtual replicas of hardware, software and firmware components, DTs enable organizations to proactively identify vulnerabilities and mitigate cyber threats with greater precision \cite{Luzzi_2024}, \cite{Holmes_2021}. While their application has demonstrated significant value in sectors such as manufacturing, construction, automotive, agriculture, and transportation, the cybersecurity implications of DT integration remain underexplored \cite{Holmes_2021}.

When embedded within a Cybersecurity Mesh architecture, DTs can improve threat prediction, streamline incident response, and improve systemic resilience \cite{McLaughlin02122023}. The integration of Large Language Models (LLMs) into this ecosystem further amplifies these benefits. LLMs can support automated interpretation of system behaviors, assist in real-time threat analysis, and generate human-readable explanations of complex security events, making the DT environment more accessible and intelligent. This synergy enables more adaptive and responsive cybersecurity operations.


Building on this perspective, the objective of this research is to explore how current technologies–––specifically Digital Twins, LLMs, and penetration testing tools–––can deepen our understanding of cyber attacks and their broader context. This approach enables a redefinition of key security functions \cite{Baldassarre2019}:

\begin{enumerate}
\item \textbf{Detection}. DTs and LLMs together redefine detection by enabling real-time simulation, analysis, and intelligent interpretation of cyber-physical environments. While DTs mirror IT, OT, and IoT infrastructures to enhance situational awareness, LLMs support the detection process through advanced natural language understanding and threat intelligence extraction from unstructured data sources, such as \emph{OSINT} and dark web forums. This synergy improves the identification of critical events and behavioral patterns across human, social, cultural, and political dimensions.

\item \textbf{Response}. DTs and LLMs enhance response strategies through dynamic simulations and real-time contextual reasoning. Digital Twins facilitate scenario testing and decision-making in crisis situations, while LLMs contribute by generating adaptive response recommendations, analyzing incident reports, and enabling more effective communication between technical teams and non-expert stakeholders. Together, they support coordinated responses across civil society, strengthening resilience against cybercrime and cyber terrorism.

\item \textbf{Prevention}. In preventive efforts, DTs and LLMs enable more nuanced risk assessment and proactive threat modeling. DTs simulate critical assets—ranging from individuals and communities to infrastructure and software—while LLMs analyze historical incident data, policy documents, and socio-political discourse to anticipate emerging threats. This integrated approach allows for a comprehensive security posture that encompasses physical, organizational, and application-level dimensions, while contextualizing cyber risks within legal, economic, and psychological frameworks.
\end{enumerate}

The paper is organized as follows: \autoref{sect:rel} describes the related works; \autoref{sect:ckc} describes the Cyber Kill Chain Model and \autoref{sect:DT_GenAI} the Logical Framework in Cyber Security Education; \autoref{sect:conc} conclude the work and explain the future developments.

\section{Related Work} \label{sect:rel}
DTs are playing an increasingly critical role in enhancing cybersecurity across complex, interconnected environments. By creating dynamic virtual replicas of physical systems—spanning IT, OT, and IoT infrastructures—DTs enable organizations to simulate, monitor, and analyze systems in real time. This capability allows for the detection of anomalies, assessment of vulnerabilities, and rehearsal of threat scenarios in a safe, controlled environment.

Integration of digital twins with intrusion detection systems has shown promise in detecting cyberattacks and understanding resource impacts in IoT-based smart city infrastructures \cite{El-Hajj_2024}. Studies emphasize the importance of addressing cybersecurity challenges in digital twin deployments, particularly in smart cities, where risks such as unauthorized access and data manipulation are significant concerns. To mitigate these risks, multi-layered security frameworks incorporating encryption, access control, and anomaly detection have been proposed. The integration of digital twins with Building Information Models (BIM) and IoT technologies is seen as crucial for enhancing cybersecurity in the built environment \cite{alshammari2021cybersecurity}. Overall, digital twins offer a promising approach to improve cyber resilience by enabling real-time monitoring and virtualization of physical systems \cite{faleiro2021digital}.

Instead, regarding LLMs, recent research explores the integration in automated penetration testing, aiming to enhance efficiency and effectiveness. Studies have demonstrated LLMs' potential in automating various aspects of penetration testing, including task planning, vulnerability hunting, and attack vector suggestion \cite{hassanin2024comprehensiveoverviewlargelanguage}. PentestGPT, an LLM-empowered tool, has shown significant improvements in task completion rates compared to traditional LLM approaches \cite{deng2024pentestgptllmempoweredautomaticpenetration}. Similarly, PentestAgent leverages multi-agent collaboration and techniques like Retrieval Augmented Generation to enhance penetration testing knowledge and automate intelligence gathering, vulnerability analysis, and exploitation stages \cite{shen2025pentestagentincorporatingllmagents}. While LLMs exhibit proficiency in specific sub-tasks, challenges remain in maintaining an integrated understanding of the overall testing scenario \cite{deng2024pentestgptllmempoweredautomaticpenetration}. Despite these limitations, the integration of LLMs in penetration testing shows promise in reducing time and resource requirements while maintaining high standards of accuracy and thoroughness.

\section{Cyber Kill Chain} \label{sect:ckc}

The Cyber Kill Chain is a model that outlines the stages of a cyber-attack, from initial reconnaissance to the final action \cite{Yadav2015TechnicalAO,Barletta_Craste}: (i) \textit{Reconnaissance}. The attacker gathers information about the target to plan the attack; (ii) \textit{Weaponization}. Malicious payloads are created to exploit the identified vulnerabilities; (iii) \textit{Delivery}. The attacker transmit the payload to the target environment; (iv) \textit{Exploitation}. Malware code is executed on the target system, exploiting vulnerabilities to gain unauthorized access; (v) \textit{Installation}. Malware or other malicious components are installed on compromised systems; (vi) \textit{Command and Control (C2)}. The attacker establishes communication channels to control compromised systems and execute their plans; (vii) \textit{Actions on Objectives}. The final phase where the attacker achieves their ultimate goal, such as data theft, system control, or disruption.
This model has been extensively adopted within the security community as a key framework for understanding and mitigating cyber-attacks. Nonetheless, Khan et al. \cite{Khan2022RealTimeCA} highlight its shortcomings, especially in confronting sophisticated and persistent threats. To overcome these limitations, they introduce an enhanced model that enables simultaneous analysis of multiple threat stages, mirroring the human cognitive process in threat evaluation. Tarnowski \cite{tarnowski2017use} also underscores the Cyber Kill Chain's role in strengthening cybersecurity, particularly in safeguarding networks. He argues that, when integrated with advanced analytics and predictive techniques, the Kill Chain becomes a powerful instrument for proactive, inside-out defense, capable of detecting and disrupting ongoing attacks.

\section{Digital Twins and Generative AI to improve Cyber Security Education} \label{sect:DT_GenAI}
Digital Twins (DTs)—virtual counterparts of physical systems—have found widespread applications across various industries. In manufacturing, they enable real-time monitoring, predictive maintenance, and simulation-based optimization \cite{AP_2024,Guo2022}; in healthcare, they support personalized medicine, precise diagnostics, and advanced treatment planning \cite{AP_2024,Latorre_2024}. Agriculture benefits from improved farm management and resource utilization \cite{AP_2024,Latorre_2024}, while the automotive and aviation sectors employ DTs for asset tracking and fault prediction \cite{AP_2024}. Urban planning, energy systems, and smart infrastructure also leverage DTs for efficient design, sustainability, and complex system management \cite{Latorre_2024,Joshi_2024}. These applications are enabled by the convergence of IoT, AI, XR, and cloud computing, which bridge the physical and digital realms \cite{AP_2024,Guo2022}.

Recently, the use of Digital Twins has extended into cybersecurity. Cyber Digital Twins (CDTs) can mirror IT and OT environments, enabling the simulation of cyberattacks and the evaluation of defense mechanisms in a risk-free, controllable setting \cite{Somma_2023,Biffl2019}. For instance, Varghese et al. \cite{Varghese_2022} modeled four types of cyberattacks—including command injection and false data injection—on DT-enabled systems, highlighting how such replicas can be exploited to test vulnerabilities in protocols like ENIP. Similarly, Dietz et al. \cite{Dietz_2020} used DTs to simulate man-in-the-middle attacks via ARP poisoning in industrial control systems, while El-Hajj \cite{El-Hajj_2024} demonstrated network-based attacks such as Hping3 floods and NMAP scans within DT testbeds.

Beyond attack simulation, DTs have proven valuable in Security Operations Center (SOC) environments. Hadar et al. \cite{Hadar_2020} introduced a cyber DT capable of automatically gathering and prioritizing security control requirements. Khan et al. \cite{Khan2022RealTimeCA} and Sasikala et al. \cite{Sasikala_2024} further showed that real-time analysis within DTs enables quicker detection and response to evolving threats.

To amplify these capabilities, the integration of Large Language Models (LLMs) into CDTs represents a promising frontier. LLMs, with their advanced natural language understanding and generation capabilities, can enhance threat analysis, automate the classification of alerts, interpret attack patterns, and generate adaptive cyber threat intelligence. 

Therefore, DTs of varying fidelity can support different stages of learning, with LLMs acting as mentors to enhance the educational experience \cite{lin2024transformingengineeringeducationusing}. In social media research, LLM-powered DTs like Y enable sophisticated simulations of user interactions and platform dynamics \cite{rossetti2024ysocialllmpoweredsocial}. LLMs are also transforming cybersecurity, improving threat detection, intelligence gathering, and anomaly detection \cite{Guven_2024}. In manufacturing, LLMs offer solutions to challenges in digital twin modeling, such as limited data and system analysis inefficiencies. A unified description-prediction-prescription framework has been proposed to integrate LLMs with digital twin technologies, potentially revolutionizing enterprise modeling and optimization \cite{yang2025leveraginglargelanguagemodels}.
When embedded within CDTs, LLMs can provide contextual, explainable insights in real time, improving diagnostics, anomaly detection, and decision-making support in SOCs.

These advancements highlight the growing importance of LLMs and DTs in various fields, from education to industry.
This convergence of DTs and LLMs paves the way for advanced applications such as: Enhanced cyber-attack detection and forensic simulation; Real-time intrusion detection in IoT and CPS environments; Automated vulnerability assessment and penetration testing; Dynamic threat intelligence generation; Continuous diagnostics and adaptive security controls.
Therefore, this study investigates the integration of the Cyber Kill Chain (CKC) model within Cyber Digital Twins augmented by LLMs. This approach aims to enable concurrent threat stage analysis, predictive modeling of attacker behavior, and intelligent response strategies—transforming static defense paradigms into adaptive, resilient cybersecurity ecosystems.

\begin{figure}
    \centering
    \includegraphics[width=0.85\textwidth]{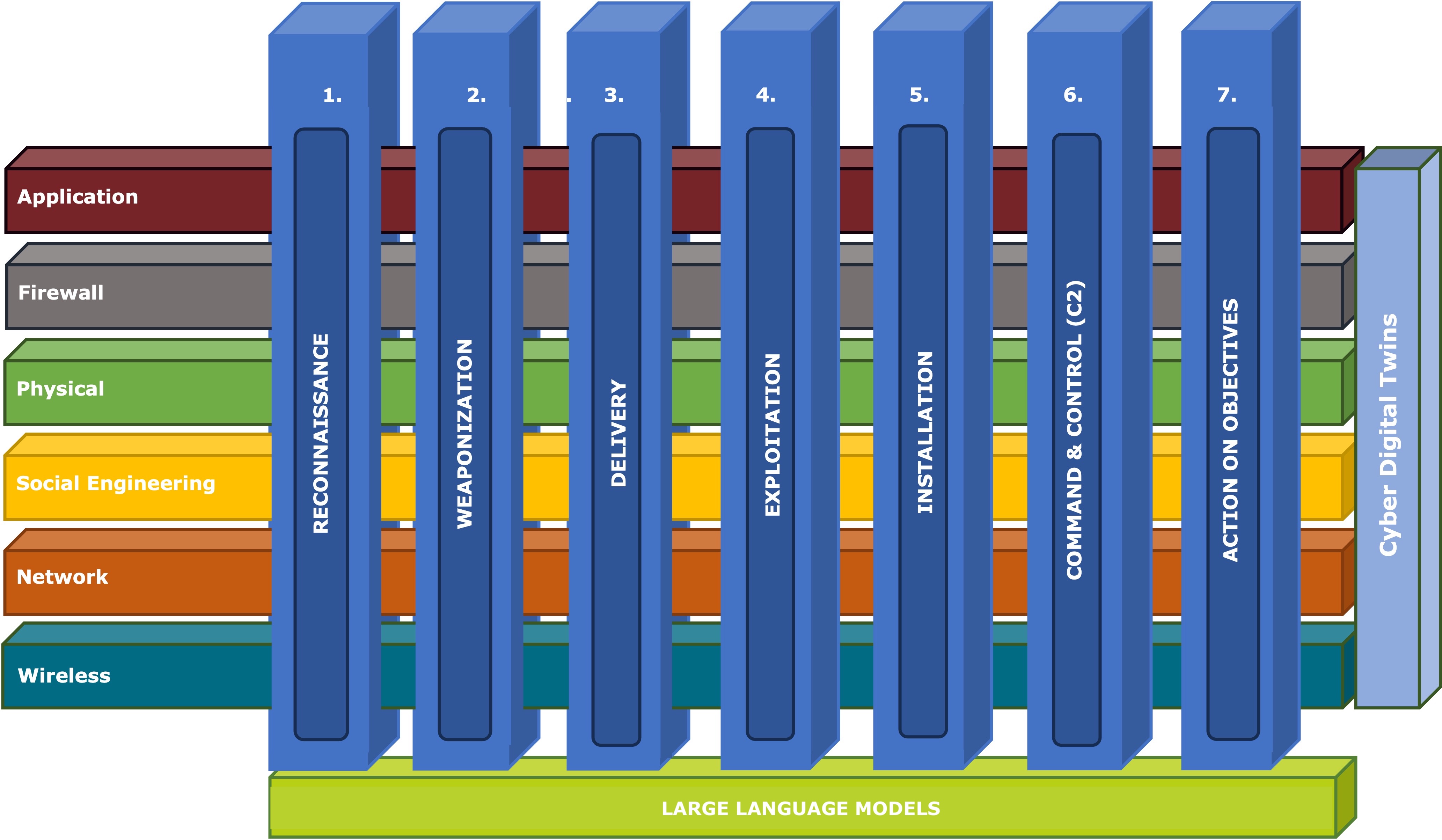}
    \caption{Logical Architecture: Cyber Digital Twin to improve Security Education} \label{RTK_CDT}
\end{figure}

Figure \ref{RTK_CDT} shows the logical architecture adopted to explain cyber attacks with LLMs through the digital twins.
The \textit{horizontal dimension} refers to the assets simulated by the Digital Twin (DT), which represent the various categories of penetration testing—namely Application, Firewall, Physical, Social Engineering, Network, and Wireless. Each of these layers enables the identification of appropriate methodologies, tools, and techniques essential for enhancing cybersecurity education in the field of Penetration Testing. In contrast, the \textit{ vertical dimension} incorporates the phases of the Cyber Kill Chain, offering a structured framework for detecting, responding to, and preventing cyberattacks.

\begin{figure}
    \centering
    \includegraphics[width=0.85\textwidth]{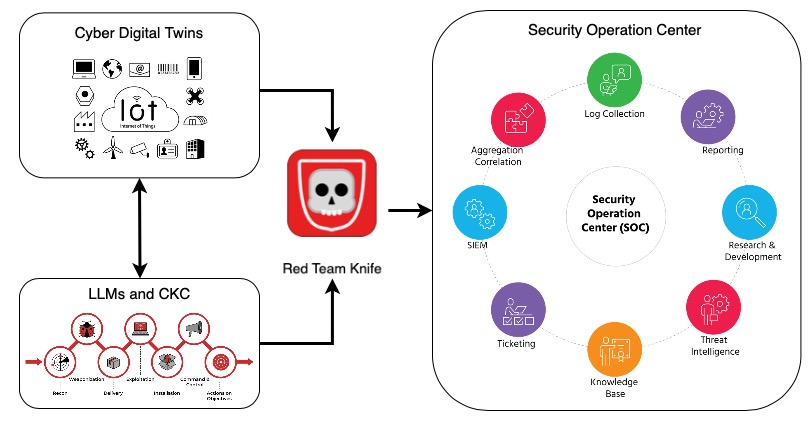}
    \caption{Red Team Knife Approach} \label{RTK_LLM}
\end{figure}

In the proposed architecture, LLMs and a Red Team Knife (RTK) play a key role for Cyber Security Education. The goal is to help perform the various activities to identify threats and vulnerabilities in the various systems through suggestions and guidelines in applying the results obtained from the various tools implemented. 
An example is shown in Figure \ref{RTK_LLM}: RTK identify vulnerabilities like SQL injection, XSS, CSRF, data exposure, etc. of the Cyber Digital Twins analyzed; thanks to the support of the "LLMs and CKC" it is possible to be able to acquire penetration testing competencies and improve the Blue Team activities like in the Secure Operation Center.

\subsection{Red team Knife}
A penetration test (or pentest) is a simulated cyberattack on a computer system, conducted to assess its security posture. The purpose of a pentest is to uncover weaknesses or vulnerabilities that could be exploited by unauthorized actors to access system functionalities or sensitive data. This process enables a thorough risk assessment of the system.

\begin{figure}
    \centering
    \includegraphics[width=0.85\textwidth]{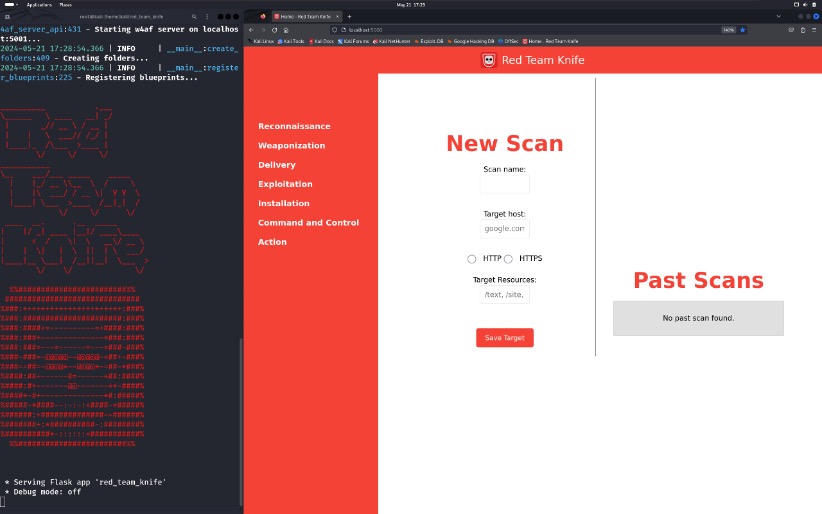}
    \caption{Red Team Knife Dashboard} \label{RTKfig1}
\end{figure}

Red Team Knife\footnote{https://github.com/Red-Team-Knife/red\_team\_knife} acts as a unified interface for a variety of widely-used red teaming tools (see Figure~\ref{RTKfig1}), such as Nmap, Nmap Vulnerability Scanner, Feroxbuster, theHarvester, Dig, w4af, an SMTP email spoofer, Commix, and Sqlmap. The main goal of Red Team Knife (RTK) is to make these tools accessible to both cybersecurity professionals and less-experienced users conducting security assessments. Despite the growing availability of penetration testing tools, there remains a gap in platforms that offer structured, comprehensive support for conducting full-spectrum security analyses. Many current solutions are either not yet mature or are limited to specific domains within cybersecurity.

\begin{figure}
    \centering
    \includegraphics[width=0.85\textwidth]{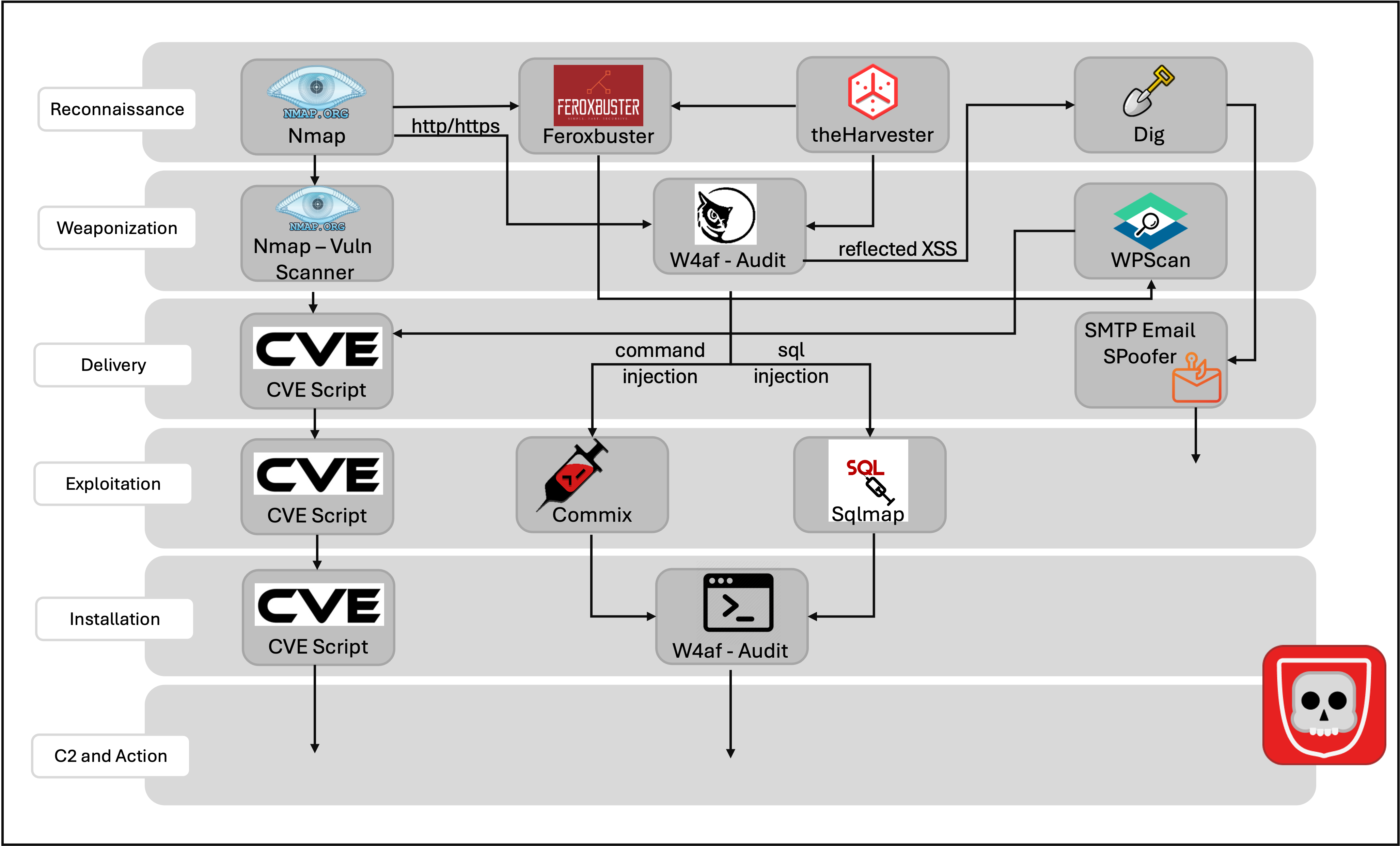}
    \caption{Cyber Kill Chain in Red Team Knife} \label{RTK_CKCfig2}
\end{figure}

RTK provides structured guidance on conducting penetration testing in line with the phases of the Cyber Kill Chain. The integrated tools are interconnected and provide contextual suggestions indicating potential next steps once a specific phase yields meaningful results.
Notably, the w3af tool may prompt the user to revisit previous stages, such as Dig, as it provides valuable insights relevant to the Reconnaissance and Weaponisation phases. This illustrates the non-linear nature of penetration testing workflows, where earlier stages can be revisited to refine or expand the attack strategy based on newly discovered information (Figure \ref{RTK_CKCfig2}).

\section{Conclusion}\label{sect:conc}
This paper explores the cybersecurity landscape and investigates how digital twins and LLMs can enhance individuals’ competencies in penetration testing. Specifically, it introduces a dual-dimensional approach to cybersecurity education: the horizontal dimension, represented by various digital twins simulating different asset types; and the vertical dimension, aligned with the stages of the Cyber Kill Chain, which together support the development of layered functionalities within the Red Team Knife (RTK) and LLMs.

Although the framework is still in its early development phase, we are currently designing user studies to assess its effectiveness and practical relevance in real-world XAI scenarios. By the time of the workshop, we anticipate sharing some preliminary insights from these studies, aiming to spark meaningful discussion and gather valuable feedback from participants.

\section*{Acknowledgements} This work was partially supported by the following projects: SERICS - ``Security and Rights In the CyberSpace - SERICS'' (PE00000014) under the MUR National Recovery and Resilience Plan funded by the European Union - NextGenerationEU; Patto territoriale "Sistema universitario pugliese" – CUP F61B23000370006.

%
%
%
\bibliographystyle{splncs04}
\bibliography{reference}

\begin{thebibliography}{10}
\providecommand{\url}[1]{\texttt{#1}}
\providecommand{\urlprefix}{URL }
\providecommand{\doi}[1]{https://doi.org/#1}

\bibitem{AP_2024}
A~P, J., Shankar, A., J~R, A.N., Koliparthy, K., Badiger, V.N., Shivakumar, V.: Exploring the applications and challenges of digital twins in various industries. In: 2024 IEEE 9th International Conference for Convergence in Technology (I2CT). pp.~1--7 (2024). \doi{10.1109/I2CT61223.2024.10543818}

\bibitem{alshammari2021cybersecurity}
Alshammari, K., Beach, T., Rezgui, Y.: Cybersecurity for digital twins in the built environment: Current research and future directions. Journal of Information Technology in Construction  \textbf{26},  159--173 (2021)

\bibitem{Baldassarre2019}
Baldassarre, M.T., Barletta, V.S., Caivano, D., Raguseo, D., Scalera, M.: Teaching cyber security: The hack-space integrated model. vol.~2315 (2019), \url{https://www.scopus.com/inward/record.uri?eid=2-s2.0-85061370504&partnerID=40&md5=e8da8bde8df7b4a276e5517e34136832}

\bibitem{Barletta_Craste}
Barletta, V.S., Caivano, D., Calvano, M., Curci, A., Piccinno, A.: Craste: Human factors and perception in cybersecurity education. vol.~3713, p. 75 – 81 (2024), \url{https://www.scopus.com/inward/record.uri?eid=2-s2.0-85198753881&partnerID=40&md5=35f9b858e583d214bb7a53c0a7dbf0da}

\bibitem{Biffl2019}
Biffl, S., Eckhart, M., L{\"u}der, A., Weippl, E.: Introduction to Security and Quality Improvement in Complex Cyber-Physical Systems Engineering, pp. 1--29. Springer International Publishing, Cham (2019)

\bibitem{deng2024pentestgptllmempoweredautomaticpenetration}
Deng, G., Liu, Y., Mayoral-Vilches, V., Liu, P., Li, Y., Xu, Y., Zhang, T., Liu, Y., Pinzger, M., Rass, S.: Pentestgpt: An llm-empowered automatic penetration testing tool (2024), \url{https://arxiv.org/abs/2308.06782}

\bibitem{Dietz_2020}
Dietz, M., Vielberth, M., Pernul, G.: Integrating digital twin security simulations in the security operations center. In: Proceedings of the 15th International Conference on Availability, Reliability and Security. ARES '20, Association for Computing Machinery, New York, NY, USA (2020)

\bibitem{El-Hajj_2024}
El-Hajj, M.: Leveraging digital twins and intrusion detection systems for enhanced security in iot-based smart city infrastructures. Electronics  \textbf{13}(19) (2024). \doi{10.3390/electronics13193941}

\bibitem{faleiro2021digital}
Faleiro, R., Pan, L., Pokhrel, S.R., Doss, R.: Digital twin for cybersecurity: Towards enhancing cyber resilience. In: International Conference on Broadband Communications, Networks and Systems. pp. 57--76. Springer (2021)

\bibitem{Guo2022}
Guo, J., Lv, Z.: Application of digital twins in multiple fields. Multimedia Tools and Applications  \textbf{81}(19),  26941--26967 (2022). \doi{10.1007/s11042-022-12536-5}, \url{https://doi.org/10.1007/s11042-022-12536-5}

\bibitem{Guven_2024}
Guven, M.: A comprehensive review of large language models in cyber security. International Journal of Computational and Experimental Science and Engineering  \textbf{10}(3) (Sep 2024). \doi{10.22399/ijcesen.469}

\bibitem{Hadar_2020}
Hadar, E., Kravchenko, D., Basovskiy, A.: Cyber digital twin simulator for automatic gathering and prioritization of security controls’ requirements. In: 2020 IEEE 28th International Requirements Engineering Conference (RE). pp. 250--259 (2020). \doi{10.1109/RE48521.2020.00035}

\bibitem{hassanin2024comprehensiveoverviewlargelanguage}
Hassanin, M., Moustafa, N.: A comprehensive overview of large language models (llms) for cyber defences: Opportunities and directions (2024), \url{https://arxiv.org/abs/2405.14487}

\bibitem{Holmes_2021}
Holmes, D., Papathanasaki, M., Maglaras, L., Ferrag, M.A., Nepal, S., Janicke, H.: Digital twins and cyber security – solution or challenge? In: 2021 6th South-East Europe Design Automation, Computer Engineering, Computer Networks and Social Media Conference (SEEDA-CECNSM). pp.~1--8 (2021). \doi{10.1109/SEEDA-CECNSM53056.2021.9566277}

\bibitem{Joshi_2024}
Joshi, R.P., Gulati, S., Kar, A.K.: Digital twin for industrial applications -- a literature review. In: Sharma, S.K., Dwivedi, Y.K., Metri, B., Lal, B., Elbanna, A. (eds.) Transfer, Diffusion and Adoption of Next-Generation Digital Technologies. pp. 92--103. Springer Nature Switzerland, Cham (2024)

\bibitem{Khan2022RealTimeCA}
Khan, M.M.S., Giraldo, J.A., Parvania, M.: Real-time cyber-physical analysis of distribution systems using digital twins. 2022 IEEE International Conference on Communications, Control, and Computing Technologies for Smart Grids (SmartGridComm) pp. 34--39 (2022), \url{https://api.semanticscholar.org/CorpusID:254153621}

\bibitem{Latorre_2024}
Latorre, L., Rego, E., De~Leo, L., Gutierrez, M.: Tech report: Digital twins (2024). \doi{https://doi.org/10.18235/0013166}

\bibitem{lin2024transformingengineeringeducationusing}
Lin, Y.Z., Alhamadah, A.H.J., Redondo, M.W., Patel, K.H., Ghimire, S., Latibari, B.S., Salehi, S., Satam, P.: Transforming engineering education using generative ai and digital twin technologies (2024), \url{https://arxiv.org/abs/2411.14433}

\bibitem{Luzzi_2024}
Luzzi, J., Naha, R., Arulappan, A., Mahanti, A.: Sok: A holistic view of cyberattacks prediction with digital twins. In: 2024 Second International Conference on Emerging Trends in Information Technology and Engineering (ICETITE). pp.~1--7 (2024). \doi{10.1109/ic-ETITE58242.2024.10493514}

\bibitem{McLaughlin02122023}
McLaughlin, K.L.: The power of digital twins in the cybersecurity mesh. EDPACS  \textbf{68}(6),  35--39 (2023). \doi{10.1080/07366981.2023.2263214}

\bibitem{rossetti2024ysocialllmpoweredsocial}
Rossetti, G., Stella, M., Cazabet, R., Abramski, K., Cau, E., Citraro, S., Failla, A., Improta, R., Morini, V., Pansanella, V.: Y social: an llm-powered social media digital twin (2024), \url{https://arxiv.org/abs/2408.00818}

\bibitem{Sasikala_2024}
Sasikala, M., Mahaboob~John, Y.M., Jothi, B., S, N., S, S.K.: Integrating digital twins with ai for real-time intrusion detection in smart infrastructure networks. In: 2024 International Conference on Intelligent Algorithms for Computational Intelligence Systems (IACIS). pp.~1--6 (2024). \doi{10.1109/IACIS61494.2024.10721892}

\bibitem{shen2025pentestagentincorporatingllmagents}
Shen, X., Wang, L., Li, Z., Chen, Y., Zhao, W., Sun, D., Wang, J., Ruan, W.: Pentestagent: Incorporating llm agents to automated penetration testing (2025), \url{https://arxiv.org/abs/2411.05185}

\bibitem{Somma_2023}
Somma, A., Casola, V., Cavalli, A.R., De~Benedictis, A., Mallouli, W., Valdés, V.E.: A cyber digital twin framework to support cyber-physical systems security. In: 2023 IEEE Smart World Congress (SWC). pp. 1--10 (2023). \doi{10.1109/SWC57546.2023.10449161}

\bibitem{tarnowski2017use}
Tarnowski, I.: How to use cyber kill chain model to build cybersecurity? European Journal of Higher Education IT  (2017)

\bibitem{Varghese_2022}
Varghese, S.A., Dehlaghi~Ghadim, A., Balador, A., Alimadadi, Z., Papadimitratos, P.: Digital twin-based intrusion detection for industrial control systems. In: 2022 IEEE International Conference on Pervasive Computing and Communications Workshops and other Affiliated Events (PerCom Workshops). p. 611–617. IEEE (Mar 2022). \doi{10.1109/percomworkshops53856.2022.9767492}

\bibitem{Yadav2015TechnicalAO}
Yadav, T.K., Rao, A.M.: Technical aspects of cyber kill chain. ArXiv  \textbf{abs/1606.03184} (2015)

\bibitem{yang2025leveraginglargelanguagemodels}
Yang, L., Luo, S., Cheng, X., Yu, L.: Leveraging large language models for enhanced digital twin modeling: Trends, methods, and challenges (2025), \url{https://arxiv.org/abs/2503.02167}

\end{thebibliography}

\end{document}